\documentclass[runningheads,orivec]{llncs}
\usepackage[T1]{fontenc}
\usepackage{graphicx,verbatim}
\usepackage{amsmath}
\usepackage{amsfonts}
\usepackage{amssymb}
\usepackage{multirow}
\usepackage{amssymb}
\usepackage{pifont}
\newcommand{\cmark}{\ding{51}}%
\newcommand{\xmark}{\ding{55}}%
\usepackage{color}
\usepackage{hyperref}

\begin{document}
\title{Pathology Image Compression with Pre-trained Autoencoders}
%

\author{
    Srikar Yellapragada\textsuperscript{1},
    Alexandros Graikos\textsuperscript{1},
    Kostas Triaridis\textsuperscript{1}, \\
    Zilinghan Li\textsuperscript{2},
    Tarak Nath Nandi\textsuperscript{2},
    Ravi K Madduri\textsuperscript{2}, \\
    Prateek Prasanna\textsuperscript{1},
    Joel Saltz\textsuperscript{1},
    Dimitris Samaras\textsuperscript{1}
}
\authorrunning{Yellapragada et al.}
\institute{
\textsuperscript{1}Stony Brook University,\ \  
\textsuperscript{2}Argonne National Laboratory 
\\
\email{myellapragad@cs.stonybrook.edu}}

\maketitle              
\begin{abstract}
The growing volume of high-resolution Whole Slide Images in digital histopathology poses significant storage, transmission, and computational efficiency challenges. Standard compression methods, such as JPEG, reduce file sizes but often fail to preserve fine-grained phenotypic details critical for downstream tasks. In this work, we repurpose autoencoders (AEs) designed for Latent Diffusion Models as an efficient learned compression framework for pathology images. We systematically benchmark three AE models with varying compression levels and evaluate their reconstruction ability using pathology foundation models. We introduce a fine-tuning strategy to further enhance reconstruction fidelity that optimizes a pathology-specific learned perceptual metric. We validate our approach on downstream tasks, including segmentation, patch classification, and multiple instance learning, showing that replacing images with AE-compressed reconstructions leads to minimal performance degradation. Additionally, we propose a K-means clustering-based quantization method for AE latents, improving storage efficiency while maintaining reconstruction quality. We provide the weights of the fine-tuned autoencoders at \href{https://huggingface.co/collections/StonyBrook-CVLab/pathology-fine-tuned-aes-67d45f223a659ff2e3402dd0}{this link}.


\end{abstract}
%
%
%

\section{Introduction}
With the increasing digitization of histopathology, large repositories of Whole Slide Images (WSIs), such as TCGA \cite{cancer2013cancer}, have been invaluable for the development of large-scale machine learning models \cite{chen2024uni,lu2024avisionlanguage,Filiot2023ScalingSSLforHistoWithMIM}. However, the size of high-resolution pathology images presents a major bottleneck in storage, transmission, and computational efficiency. A large pathology center produces over 1 million digital slides per year \cite{schuffler2021integrated}, which translates to petabytes of storage. Long-term retention of these slides, even in deep archival storage, could cost up to \$100,000 per year. As the volume of pathology data grows, developing efficient compression techniques that reduce storage requirements while preserving information relevant for AI model `consumption' remains a key issue. 

\begin{figure}[ht]
\centering
\includegraphics[width=\textwidth]{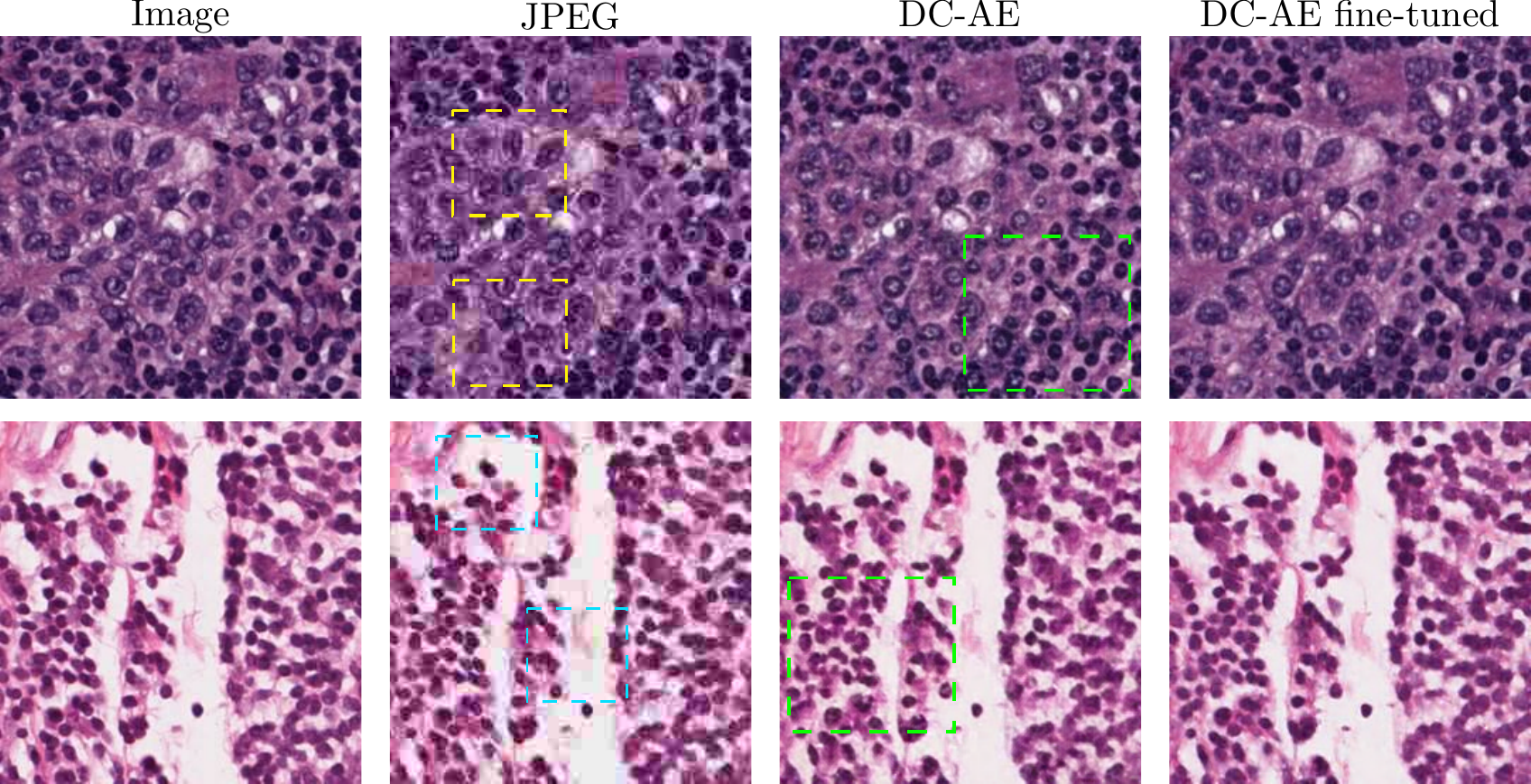}
\caption{Examples of image reconstruction using JPEG, the vanilla DC-AE \cite{chen2024deep} and fine-tuned DC-AE. JPEG at quality 10, with a comparable file size to DC-AE, introduces severe compression artifacts, including deformed nuclei and blocky artifacts (highlighted in yellow and teal). Vanilla DC-AE fails to retain certain cell structures (green), which are largely recovered through our fine-tuning strategy.}
\label{fig:reconstruction}
\vspace{-0.4cm}
\end{figure}

\begin{figure}[ht]
\centering
\includegraphics[width=1.\linewidth]{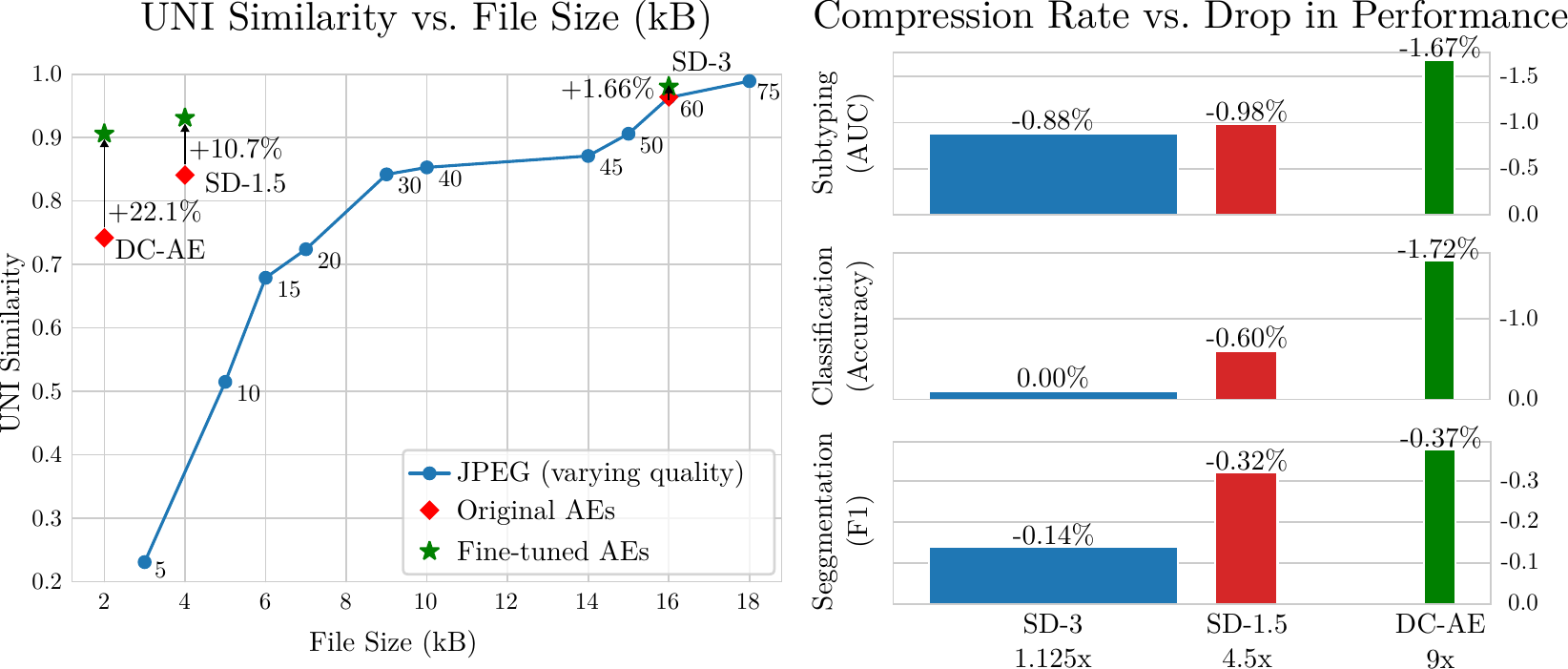}
\caption{Left: Pre-trained autoencoders outperform JPEG in reconstruction fidelity,  further improved by fine-tuning with a pathology-specific perceptual loss. Right: Using fine-tuned AE-compressed reconstructions results in minimal performance degradation. The width of each bar denotes the relative sizes of the compressed representation.}
\label{fig:teaser}
\vspace{-0.4cm}
\end{figure}

Looking for ways to reduce image size, generative diffusion models \cite{ho2020denoising} have employed autoencoders (AEs) for efficient high-resolution image synthesis. The Latent Diffusion Model (LDM) framework \cite{rombach2022high} introduced a trained autoencoder that first compresses images into a lower-dimensional latent representation before applying the diffusion model. These autoencoders are designed to maintain high reconstruction fidelity while preserving spatial locality and ensuring generalizability. While originally optimized for generative modeling, they are equally well-suited for image compression \cite{tellez2019neural,fischer2023enhanced}, making pre-trained LDM autoencoders a promising choice across diverse image domains and sizes.

In this work, we repurpose LDM autoencoders as efficient image compression models for histopathology. Conventional compression methods, such as JPEG, can significantly reduce image sizes but struggle to preserve the fine-grained pathology features critical for downstream tasks when aggressively compressing \cite{chen2020quantitative}. By leveraging the learned representations of LDM autoencoders, we achieve high compression rates while maintaining the essential pathology details. As shown in Fig.~\ref{fig:reconstruction}, heavy JPEG compression leads to substantial artifacts, whereas comparable compression with the DC-AE autoencoder model gives coherent reconstructions.

We systematically benchmark three AEs — Stable Diffusion 1.5 (SD-1.5) \cite{rombach2022high}, Stable Diffusion 3 (SD-3) \cite{esser2024scaling}, and Deep Compression Autoencoder (DC-AE-f32) \cite{chen2024deep} -- each offering different compression rates. We assess the perceptual similarity between original and reconstructed images using pathology foundation models \cite{chen2024uni,filiot2024phikon,xu2024whole}, finding that existing AEs perform surprisingly well (Fig.\ref{fig:teaser} left).
To further improve the AE reconstructions, we propose a fine-tuning strategy that optimizes the decoder for a pathology-specific learned perceptual metric. This fine-tuning strategy further aligns pathology-specific features between the reconstructions and the original without altering the desirable properties (generalizability, locality) of the autoencoder.

Beyond perceptual similarity, we validate our compression pipeline on multiple downstream tasks, including segmentation, patch and multiple instance learning classification. We show that replacing raw images with AE-compressed reconstructions results in minimal performance degradation, demonstrating the practical viability of our approach (Fig.\ref{fig:teaser} right). Finally, we introduce a K-means clustering-based quantization for the AE latents that considers the unique characteristics of the latent distribution. By mapping the continuous latent representations to a fixed set of discrete values, we further reduce storage while preserving high reconstruction fidelity, outperforming static 8-bit (int8) rounding-based quantization, which introduces artifacts and degrades image quality.

Our contributions are as follows
\begin{itemize}
    \item We repurpose LDM autoencoders for pathology image compression, demonstrating their ability to achieve high compression rates while preserving essential phenotypical details.
    \item We benchmark three AEs across multiple compression rates and evaluate their performance using pathology foundation models.
    \item  We introduce a fine-tuning strategy to enhance reconstruction fidelity by aligning AE latents with a pathology-specific learned perceptual metric.
    \item We enhance storage efficiency with a K-means clustering-based quantization for AE latents, outperforming static int8 quantization, while preserving reconstruction fidelity.
\end{itemize}


\section{Related Work}
\textbf{Autoencoders for compression.} Using autoencoders for data compression has been extensively studied with previous works showing that learned compression schemes can outperform standard compression algorithms \cite{theis2017lossy,balle2017end}. However, only recently have autoencoders trained on large-scale datasets, such as those in LDMs, been explored for image compression.

\textbf{LDM autoencoders.} The autoencoders used in LDMs embed the images into a structured, 2D latent space that downscales the image size. The AEs are trained with reconstruction, perceptual \cite{zhang2018unreasonable} and patch-based adversarial losses \cite{isola2017image} to ensure the high fidelity of the outputs. To constrain the learned latent space, Stable Diffusion \cite{rombach2022high} employs KL regularization \cite{kingma2014autoencoding} that centers the latent embeddings around zero and imposes unit variance. In DC-AE \cite{chen2024deep} there is no explicit regularization on the learned latent space, but it still maintains similar properties to the latent spaces learned by the SD AEs.

\textbf{Image compression in digital pathology.} The current go-to approach for image compression in digital pathology is JPEG \cite{fischer2025unlocking}. Previous attempts at learned compression for pathology images have also used autoencoder models \cite{tellez2019neural,fischer2023enhanced} with recent works proposing domain-specific decompositions for increased efficiency \cite{fischer2024learned}. However, all existing learning-based compression schemes fail to generalize outside the training data distribution \cite{fischer2025unlocking}. In contrast, the LDM autoencoders have been designed and trained to work with web images, allowing them to be used in a wider array of settings.

\section{Pathology image compression with AEs}
\subsection{Compression ratio}
The starting point is the comparison of storage requirements and reconstruction fidelity between JPEG compression, with different quality settings, and the learned autoencoder compression schemes. We utilize 1000 256 $\times$ 256 px image patches from the TCGA dataset to measure reconstruction fidelity. For the autoencoders we choose SD-1.5 \cite{rombach2022high}, SD-3 \cite{esser2024scaling} and DC-AE \cite{chen2024deep}. The SD-1.5 and SD-3 autoencoders downsample the image by a factor of 8, with SD-1.5 using 4 channels in the embedding whereas SD-3 uses 16 channels. The DC-AE f32 variant downsamples the image by a factor of 32 and uses 32 latent channels.

In Table~\ref{tab:compression}, we compare the compression rates of JPEG at different quality settings against the autoencoder models. To assess reconstruction fidelity, we employ three pathology foundation models—UNI \cite{chen2024uni}, Phicon-v2 \cite{filiot2024phikon}, and Gigapath \cite{xu2024whole}. Foundation pathology models \cite{chen2024uni,filiot2024phikon} train an image encoder that embeds images into a learned embedding vector in $\mathbb{R}^k$. These encoders maximize the similarity of the k-dimensional embeddings for images that are semantically and visually similar while minimizing it for different images.

In our assessment, we extract embeddings from both the original and reconstructed images and compute their cosine similarity. Traditional image quality metrics, such as SSIM and PSNR  capture pixel-level differences and are inadequate for accurately assessing the quality of pathology images \cite{xu2024histo}. Embedding similarity however provides a more task-relevant evaluation, as it correlates better with downstream performance. Our results show that existing autoencoders already achieve adequate reconstruction fidelity; even the aggressive compression performed by DC-AE maintains better embedding similarity than JPEG while requiring half the storage. 

\begin{table}[ht]
\centering
\caption{Compression metrics for JPEG and different LDM autoencoders. Fine-tuning boosts reconstruction fidelity. Employing a K-means-based quantization allows for minimal storage requirements while preserving important pathology image features.}
\begin{tabular}{c|c|c|c|ccc|cc}
\hline
\multirow{2}{*}{Compression}      & \multirow{2}{*}{Fine-tuned} & \multirow{2}{*}{Quant} & \multirow{2}{*}{\begin{tabular}[c]{@{}c@{}}Size\\  (KB)\end{tabular}} & \multicolumn{3}{c|}{Embedding similarity} & \multicolumn{2}{c}{Image quality} \\
                          &                      &              &           & UNI          & Phicon-v2    & Gigapath    & SSIM            & PSNR            \\ \hline
JPEG - 75                 & $\boldsymbol{-}$           & $\boldsymbol{-}$           & 18        & 0.988     & 0.985     & 0.985    & 0.994        & 41.93       \\
JPEG - 50                 & $\boldsymbol{-}$                   & $\boldsymbol{-}$           & 15        & 0.904     & 0.855     & 0.866    & 0.964        & 32.97       \\
JPEG - 20                 & $\boldsymbol{-}$                   & $\boldsymbol{-}$           & 7         & 0.734     & 0.623     & 0.645    & 0.870        & 27.69       \\
JPEG - 10                 & $\boldsymbol{-}$                   & $\boldsymbol{-}$           & 5         & 0.512     & 0.456     & 0.407    & 0.788        & 25.13       \\ \hline
\multirow{4}{*}{SD-1.5}   & \xmark                   & \xmark        & 16        & 0.837     & 0.825     & 0.796    & 0.651        & 22.55       \\
                          & \cmark & \xmark        & 16        & 0.932     & 0.935     & 0.909    & 0.649        & 22.50       \\
                          &  \cmark                    & static-int8   & 4         & 0.912     & 0.921     & 0.878    & 0.639        & 21.99       \\
                          &  \cmark                    & K-means      & 4         & 0.932     & 0.935     & 0.909    & 0.649        & 22.49       \\ \hline
\multirow{4}{*}{SD-3}     & \xmark                   & \xmark        & 64        & 0.959     & 0.931     & 0.947    & 0.894        & 28.25       \\
                          & \cmark & \xmark        & 64        & 0.978     & 0.967     & 0.972    & 0.877        & 27.42       \\
                          &   \cmark                   & static-int8   & 16        & 0.944     & 0.934     & 0.932   & 0.862        & 26.67       \\
                          &   \cmark                   & K-means      & 16        & 0.978     & 0.967     & 0.972    & 0.877        & 27.41       \\ \hline
\multirow{4}{*}{DC-AE-f32} & \xmark                   & \xmark        & 8         & 0.733     & 0.747     & 0.659    & 0.536        & 20.82       \\
                          & \cmark & \xmark        & 8         & 0.906     & 0.925     & 0.868    & 0.538        & 21.03       \\
                          &   \cmark                   & static-int8   & 2         & 0.900     & 0.921     & 0.861    & 0.537        & 21.06       \\
                          &   \cmark                   & K-means      & 2         & 0.906     & 0.926     & 0.867    & 0.538        & 21.03       \\ \hline
\end{tabular}
\label{tab:compression}
\end{table}

\vspace{-1cm}
\subsection{Pathology fine-tuning for AEs}
Although existing autoencoders produce faithful reconstructions, they also make non-negligible changes to specific pathology features that may be critical in downstream tasks. In Fig.~\ref{fig:reconstruction} we demonstrate one such example, where the DC-AE model changes the cell contents and structures in its reconstructions. We develop a simple fine-tuning scheme for pre-trained autoencoders to better align the image reconstructions, retaining important pathology features.

To perform this alignment, we utilize foundation pathology models and fine-tune only the \textbf{decoder} with an additional loss that maximizes the similarity of the reconstruction and the original image in the learned embedding space of a foundation model. Specifically, for an image $x$ and its reconstruction $y$, we use UNI \cite{chen2024uni} as the foundation model and minimize the L1 distance between the embeddings produced by the UNI encoder for them -
\begin{equation}
    \mathcal{L}_{\text{UNI}}(x,y) = ||\text{UNI}(x) - \text{UNI}(y)||_1.
    \label{eq:uni_loss}
\end{equation}

Following the training scheme of previous autoencoders \cite{rombach2022high}, we use an L1 pixel reconstruction loss and a learned PatchGAN discriminator loss \cite{isola2017image}. We select 2500 WSIs from TCGA Breast, Colon, and Prostate. Using DSMIL  \cite{li2021dual}, we extract 256 $\times$ 256 patches at 20$\times$ magnification, yielding 24 million patches. We utilize the codebase of \cite{rombach2022high} to finetune the autoencoders.  We set the learning rate at $5 \times 10^{-5}$ with a warmup of 10,000 steps. We train for 120,000 iterations on 8 NVIDIA A100 GPUs, with a batch size of 12 per GPU. The loss used is
\begin{equation}
    \mathcal{L}(x,y) = w_{\text{L1}}\mathcal{L}_{\text{L1}}(x,y)
    + w_{\text{GAN}}\mathcal{L}_{\text{GAN}}(x,y)
    + w_{\text{UNI}}\mathcal{L}_{\text{UNI}}(x,y)
    \label{eq:vae_finetune_loss}
\end{equation}
where we choose $w_{\text{L1}}=1$, $w_{\text{GAN}}=0.5$ and $w_{\text{UNI}}=1$.

We find that training only the decoder is enough to improve the pathology-specific reconstruction metrics, even when using a relatively small dataset. By preserving the encoder, we ensure that the AE latent space is unchanged and we only learn to interpret it differently during reconstruction. We validate the assumption that the existing autoencoders can already compress pathology images adequately and that we only need to slightly alter the reconstruction part. Fine-tuning the full autoencoder model would require significantly more data to ensure that the learned latent would not overfit to the fine-tuning dataset.

The results presented in Table~\ref{tab:compression} show that fine-tuning leads to a significant boost in embedding similarity across all models. Notably, DC-AE improves from 0.733 to 0.906, demonstrating that fine-tuning enables even highly compressed representations to retain critical information. Additionally, fine-tuning enables AEs to surpass JPEG in embedding similarity at comparable compression ratios. For example, fine-tuned SD-1.5 achieves a UNI similarity of 0.932, whereas JPEG-50 -- despite having a similar compression ratio -- only reaches 0.904, indicating that learned compression better preserves meaningful information.

\subsection{Quantization}
To further reduce storage, we apply quantization to the AE latents, converting the continuous-valued representations into more compact discrete values. The straightforward approach is static int8 quantization, where the range of the latent representations is split into equally spaced bins and mapped to 8-bit integers. However, this method does not consider the statistics of the latent representations; we find that the values are not uniformly distributed but have higher concentration near 0, which can be partially attributed to the regularization applied to the AEs (e.g. KL). This naive approach noticeably degrades reconstruction quality due to the misalignment between chosen quantization bins and the actual latent distribution.

To mitigate this, we propose a clustering-based quantization strategy that adapts to the distribution of latent values. Instead of mapping values to fixed bins, we first learn a set of representative centroids using K-means clustering and then quantize the latents based on these learned clusters. The process consists of the following steps:
(1) \textbf{Cluster learning}: We extract latents from randomly sampled TCGA images and apply K-means clustering on the values, learning 256 clusters (8-bit).
(2) \textbf{Compression}: Each latent value is assigned to its closest cluster center, only storing the cluster index instead of the floating-point value.
(3) \textbf{Decompression}: Stored indices are replaced with their respective float-valued cluster centers, followed by passing the latent through the AE decoder.


Table~\ref{tab:compression} shows that static int8 quantization leads to a non-trivial drop in embedding similarity across all models. In contrast, K-means clustering-based quantization attains nearly the same embedding similarity as the non-quantized representations while considerably reducing storage requirements.

\section{Downstream tasks with autoencoder reconstructions}
We benchmark the downstream performance of all 3 pre-trained and fine-tuned AEs on dense pixel-level and image-level tasks. For those benchmarks, we encode the images into latents using the AEs and reconstruct the image from the compressed representations. We compare the performance of the downstream methods using the original images vs using the reconstructed images.

\subsection{Image-level tasks}
For slide-level classification on TCGA-BRCA, we perform subtyping (Invasive Ductal Carcinoma vs. Invasive Lobular Carcinoma) using ABMIL \cite{ilse2018attention}. For patch-level classification, we classify images from the NCT-CRC dataset \cite{kather_2018_1214456} into nine tissue classes. In both cases, we perform 10-fold cross validation.

Table~\ref{tab:classification} shows that existing AE reconstructions already achieve strong classification performance, with minimal drop compared to original images. Fine-tuning further closes the gap, particularly for SD-1.5 and DC-AE-f32, bringing them closer to the original image results. This highlights the effectiveness of our approach and suggests that high representation similarity in the foundation model embedding space translates to functional similarity in predictive tasks.

\begin{table}[ht]
\centering
\caption{Classification using original (uncompressed) images and AE-reconstructed images. Fine-tuning improves performance, bringing it closer to that of original images.}
\begin{tabular}{c|c|cc}
\hline
Compression               & Fine-tuned & BRCA subtyping    & NCT-CRC          \\ \hline
Original images           & $\boldsymbol{-}$ & 94.89 $\pm$ 2.67  & 96.31 $\pm$ 0.20 \\ \hline
\multirow{2}{*}{SD-1.5}   & \xmark     & 92.89 $\pm$ 2.51  & 94.36 $\pm$ 0.48 \\
                          & \cmark     & 93.96 $\pm$ 2.77  & 95.73 $\pm$ 0.21 \\ \hline
\multirow{2}{*}{SD-3}     & \xmark     & 94.44 $\pm$ 2.65  & 95.99 $\pm$ 0.16 \\
                          & \cmark     & 94.05 $\pm$ 3.40  & 96.41 $\pm$ 0.14 \\ \hline
\multirow{2}{*}{DC-AE-f32} & \xmark     & 92.82 $\pm$ 1.81  & 89.23 $\pm$ 0.61 \\
                          & \cmark     & 93.30 $\pm$ 2.15  & 94.65 $\pm$ 0.29 \\ \hline
\end{tabular}
\label{tab:classification}
\end{table}

\subsection{Pixel-level tasks}
We perform segmentation using SAM-path \cite{zhang2023sam} on two datasets -- the Breast Cancer Semantic Segmentation (BCSS) dataset \cite{amgad2019bcss} and the Colorectal Adenocarcinoma Gland (CRAG) dataset \cite{GRAHAM2019199crag}. BCSS contains patches sampled from 151 TCGA-BRCA WSIs on which we perform tissue region segmentation. CRAG contains patches sampled from 38 colon cancer WSIs, and we perform semantic segmentation of colorectal adenocarcinoma and benign glands. 

In Table~\ref{tab:segmentation} we showcase the average Dice score, intersection over union (IoU) and F1 score for the original images, JPEG-compressed images and autoencoder reconstructions. For the BCSS dataset, we perform comparable compression to JPEG-10 (32 vs 57 KB) when using our K-means quantization, while only having <1\% performance drop for all metrics. With JPEG-10 the performance decreases by >10\%. Similarly, on CRAG we apply similar compression (72 vs 93 KB) without altering the result while JPEG-10 reduces the Dice score by >3\%.


\begin{table}[ht]
\centering
\caption{Segmentation results using original, JPEG and AE-compressed images. The original CRAG images are uncompressed whereas BCSS are compressed with JPEG-75. Heavy JPEG compression corrupts task-important features while AE compression (with and without fine-tuning) with quantization, does not sacrifice performance.} 
\begin{tabular}{c|c|c|c|ccc|c|ccc}
\hline
\multirow{2}{*}{Compression}                       & \multirow{2}{*}{FT} & \multirow{2}{*}{Quant} & \multicolumn{4}{c|}{BCSS}    & \multicolumn{4}{c}{CRAG}   \\ \cline{4-11}
                       &          &           & Size (KB) & Dice $\uparrow$    & F1 $\uparrow$   & IoU $\uparrow$  & Size (KB)& Dice $\uparrow$   & F1 $\uparrow$   & IoU $\uparrow$  \\ 
                       \hline
Original images           & $\boldsymbol{-}$       & $\boldsymbol{-}$              & 545& 71.57   & 80.38 & 67.20 &3935& 87.16   & 85.24 & 85.24 \\ 
\hline
JPEG 50           & $\boldsymbol{-}$       & $\boldsymbol{-}$              &154  &  71.16       &   79.92
   &   66.56&  246  &  86.87  & 85.38 & 85.38 \\ 

JPEG 10           & $\boldsymbol{-}$       & $\boldsymbol{-}$              & 57 &   60.56      &   68.02
     &   51.54&  93  &   83.45   & 84.87      &  84.87    \\ 
\hline
\multirow{3}{*}{SD-1.5}   & \xmark & \xmark         & 256& 71.34   & 80.12 &66.84 & 576 &86.57  & 84.34 & 84.34 \\
                          & \cmark & \xmark          &256& 71.41   & 80.20 & 66.95 & 576 &86.80& 85.48 & 85.48 \\ 
                          & \cmark & \cmark          &64 & 71.34   & 80.13 & 66.84& 144& 86.82   & 85.50 & 85.50 \\ \hline
\multirow{3}{*}{SD-3}     & \xmark & \xmark          &512& 71.52   & 80.33 & 67.12& 1152&  86.85  & 84.49 & 84.49 \\
                          & \cmark & \xmark          &512 & 71.46   & 80.26 & 67.02&1152 &  86.90  & 84.80  &  84.80\\ 
                          & \cmark & \cmark          & 128& 71.47   & 80.27 & 67.04& 288&  86.90  & 84.81  &  84.81\\ \hline
\multirow{3}{*}{DC-AE-f32} & \xmark & \xmark          &128&71.25   & 80.02 & 66.70 &288&  87.27  & 87.62 & 87.62 \\
                          & \cmark & \xmark          &128 &71.21   & 79.97 & 66.63 &288&  87.20  & 88.06 & 88.06 \\ 
                          & \cmark & \cmark          &32&71.20   & 79.96 & 66.62 &72 &87.20  & 88.08 & 88.08 \\ \hline
\end{tabular}

\label{tab:segmentation}
\end{table}

\vspace{-1cm}
\section{Conclusion}
In this work, we propose a new digital histopathology image compression scheme using pre-trained autoencoder models. We find that autoencoders trained for LDMs can effectively compress pathology images better than the current widely-used compression algorithms, such as JPEG. Furthermore, we showed that we can further improve the autoencoder reconstructions with small-scale fine-tuning using a pathology-specific perceptual metric.

A limitation of our method is that decompression using AEs is slower than JPEG, which may impact real-time applications. Despite this limitation, we believe that our work can significantly impact the digital histopathology field, where data storage remains a significant issue. By developing better compression schemes we can increase the data availability, which is necessary for future foundation model training in the pathology domain.

%
%
%
\bibliographystyle{splncs04}
\bibliography{main}

\begin{thebibliography}{10}
\providecommand{\url}[1]{\texttt{#1}}
\providecommand{\urlprefix}{URL }
\providecommand{\doi}[1]{https://doi.org/#1}

\bibitem{amgad2019bcss}
Amgad, M., Elfandy, H., Hussein, H., Atteya, L.A., Elsebaie, M.A., Abo~Elnasr,
  L.S., Sakr, R.A., Salem, H.S., Ismail, A.F., Saad, A.M., et~al.: Structured
  crowdsourcing enables convolutional segmentation of histology images.
  Bioinformatics  \textbf{35}(18),  3461--3467 (2019)

\bibitem{balle2017end}
Ball{\'e}, J., Laparra, V., Simoncelli, E.P.: End-to-end optimized image
  compression. In: 5th International Conference on Learning Representations,
  ICLR 2017 (2017)

\bibitem{cancer2013cancer}
Cancer Genome Atlas Research~Network, J., et~al.: The cancer genome atlas
  pan-cancer analysis project. Nat. Genet  \textbf{45}(10),  1113--1120 (2013)

\bibitem{chen2024deep}
Chen, J., Cai, H., Chen, J., Xie, E., Yang, S., Tang, H., Li, M., Lu, Y., Han,
  S.: Deep compression autoencoder for efficient high-resolution diffusion
  models. arXiv preprint arXiv:2410.10733  (2024)

\bibitem{chen2024uni}
Chen, R.J., Ding, T., Lu, M.Y., Williamson, D.F., Jaume, G., Chen, B., Zhang,
  A., Shao, D., Song, A.H., Shaban, M., et~al.: Towards a general-purpose
  foundation model for computational pathology. Nature Medicine  (2024)

\bibitem{chen2020quantitative}
Chen, Y., Janowczyk, A., Madabhushi, A.: Quantitative assessment of the effects
  of compression on deep learning in digital pathology image analysis. JCO
  clinical cancer informatics  \textbf{4},  221--233 (2020)

\bibitem{esser2024scaling}
Esser, P., Kulal, S., Blattmann, A., Entezari, R., M{\"u}ller, J., Saini, H.,
  Levi, Y., Lorenz, D., Sauer, A., Boesel, F., et~al.: Scaling rectified flow
  transformers for high-resolution image synthesis. In: Forty-first
  international conference on machine learning (2024)

\bibitem{Filiot2023ScalingSSLforHistoWithMIM}
Filiot, A., Ghermi, R., Olivier, A., Jacob, P., Fidon, L., Kain, A.M.,
  Saillard, C., Schiratti, J.B.: Scaling self-supervised learning for
  histopathology with masked image modeling. medRxiv  (2023).
  \doi{10.1101/2023.07.21.23292757},
  \url{https://www.medrxiv.org/content/early/2023/07/26/2023.07.21.23292757}

\bibitem{filiot2024phikon}
Filiot, A., Jacob, P., Mac~Kain, A., Saillard, C.: Phikon-v2, a large and
  public feature extractor for biomarker prediction. arXiv preprint
  arXiv:2409.09173  (2024)

\bibitem{fischer2024learned}
Fischer, M., Maier-Hein, K.: Learned image compression for he-stained
  histopathological images via stain deconvolution. In: Medical Optical Imaging
  and Virtual Microscopy Image Analysis: Second International Workshop, MOVI
  2024, Held in Conjunction with MICCAI 2024, Marrakesh, Morocco, October 10,
  2024, Proceedings. p.~97. Springer Nature (2024)

\bibitem{fischer2023enhanced}
Fischer, M., Neher, P., Sch{\"u}ffler, P., Xiao, S., Almeida, S.D., Ulrich, C.,
  Muckenhuber, A., Braren, R., G{\"o}tz, M., Kleesiek, J., et~al.: Enhanced
  diagnostic fidelity in pathology whole slide image compression via deep
  learning. In: International Workshop on Machine Learning in Medical Imaging.
  pp. 427--436. Springer (2023)

\bibitem{fischer2025unlocking}
Fischer, M., Neher, P., Sch{\"u}ffler, P., Ziegler, S., Xiao, S., Peretzke, R.,
  Clunie, D., Ulrich, C., Baumgartner, M., Muckenhuber, A., et~al.: Unlocking
  the potential of digital pathology: Novel baselines for compression. Journal
  of Pathology Informatics p. 100421 (2025)

\bibitem{GRAHAM2019199crag}
Graham, S., Chen, H., Gamper, J., Dou, Q., Heng, P.A., Snead, D., Tsang, Y.W.,
  Rajpoot, N.: Mild-net: Minimal information loss dilated network for gland
  instance segmentation in colon histology images. Medical Image Analysis
  \textbf{52},  199--211 (2019).
  \doi{https://doi.org/10.1016/j.media.2018.12.001},
  \url{https://www.sciencedirect.com/science/article/pii/S1361841518306030}

\bibitem{ho2020denoising}
Ho, J., Jain, A., Abbeel, P.: Denoising diffusion probabilistic models.
  Advances in neural information processing systems  \textbf{33},  6840--6851
  (2020)

\bibitem{ilse2018attention}
Ilse, M., Tomczak, J., Welling, M.: Attention-based deep multiple instance
  learning. In: International conference on machine learning. pp. 2127--2136.
  PMLR (2018)

\bibitem{isola2017image}
Isola, P., Zhu, J.Y., Zhou, T., Efros, A.A.: Image-to-image translation with
  conditional adversarial networks. In: Proceedings of the IEEE conference on
  computer vision and pattern recognition. pp. 1125--1134 (2017)

\bibitem{kather_2018_1214456}
Kather, J.N., Halama, N., Marx, A.: {100,000 histological images of human
  colorectal cancer and healthy tissue} (May 2018).
  \doi{10.5281/zenodo.1214456}, \url{https://doi.org/10.5281/zenodo.1214456}

\bibitem{kingma2014autoencoding}
Kingma, D.P., Welling, M.: Auto-encoding variational bayes. In: Bengio, Y.,
  LeCun, Y. (eds.) 2nd International Conference on Learning Representations,
  {ICLR} 2014, Banff, AB, Canada, April 14-16, 2014, Conference Track
  Proceedings (2014), \url{http://arxiv.org/abs/1312.6114}

\bibitem{li2021dual}
Li, B., Li, Y., Eliceiri, K.W.: Dual-stream multiple instance learning network
  for whole slide image classification with self-supervised contrastive
  learning. In: Proceedings of the IEEE/CVF conference on computer vision and
  pattern recognition. pp. 14318--14328 (2021)

\bibitem{lu2024avisionlanguage}
Lu, M.Y., Chen, B., Williamson, D.F., Chen, R.J., Liang, I., Ding, T., Jaume,
  G., Odintsov, I., Le, L.P., Gerber, G., et~al.: A visual-language foundation
  model for computational pathology. Nature Medicine  \textbf{30},  863–874
  (2024)

\bibitem{rombach2022high}
Rombach, R., Blattmann, A., Lorenz, D., Esser, P., Ommer, B.: High-resolution
  image synthesis with latent diffusion models. In: Proceedings of the IEEE/CVF
  conference on computer vision and pattern recognition. pp. 10684--10695
  (2022)

\bibitem{schuffler2021integrated}
Sch{\"u}ffler, P.J., Geneslaw, L., Yarlagadda, D.V.K., Hanna, M.G., Samboy, J.,
  Stamelos, E., Vanderbilt, C., Philip, J., Jean, M.H., Corsale, L., et~al.:
  Integrated digital pathology at scale: a solution for clinical diagnostics
  and cancer research at a large academic medical center. Journal of the
  American Medical Informatics Association  \textbf{28}(9),  1874--1884 (2021)

\bibitem{tellez2019neural}
Tellez, D., Litjens, G., Van~der Laak, J., Ciompi, F.: Neural image compression
  for gigapixel histopathology image analysis. IEEE transactions on pattern
  analysis and machine intelligence  \textbf{43}(2),  567--578 (2019)

\bibitem{theis2017lossy}
Theis, L., Shi, W., Cunningham, A., Husz{\'a}r, F.: Lossy image compression
  with compressive autoencoders. In: International Conference on Learning
  Representations (2017)

\bibitem{xu2024whole}
Xu, H., Usuyama, N., Bagga, J., Zhang, S., Rao, R., Naumann, T., Wong, C.,
  Gero, Z., Gonz{\'a}lez, J., Gu, Y., et~al.: A whole-slide foundation model
  for digital pathology from real-world data. Nature  \textbf{630}(8015),
  181--188 (2024)

\bibitem{xu2024histo}
Xu, X., Kapse, S., Prasanna, P.: Histo-diffusion: A diffusion super-resolution
  method for digital pathology with comprehensive quality assessment. arXiv
  preprint arXiv:2408.15218  (2024)

\bibitem{zhang2023sam}
Zhang, J., Ma, K., Kapse, S., Saltz, J., Vakalopoulou, M., Prasanna, P.,
  Samaras, D.: Sam-path: A segment anything model for semantic segmentation in
  digital pathology. In: International Conference on Medical Image Computing
  and Computer-Assisted Intervention. pp. 161--170. Springer (2023)

\bibitem{zhang2018unreasonable}
Zhang, R., Isola, P., Efros, A.A., Shechtman, E., Wang, O.: The unreasonable
  effectiveness of deep features as a perceptual metric. In: Proceedings of the
  IEEE conference on computer vision and pattern recognition. pp. 586--595
  (2018)

\end{thebibliography}

\end{document}